\documentclass{llncs}

\usepackage{algorithm}
\usepackage[noend]{algpseudocode}
\usepackage{color}
\usepackage{cprotect}
\usepackage{flushend}
\usepackage{graphicx}
\usepackage{multirow}
\usepackage{nohyperref}
\usepackage[final]{listings}
\usepackage{textcomp}
\usepackage{tikz}
\usepackage{url}
\usepackage{verbatim}
\usepackage{xspace}

\newcommand{\ie}{{\em i.e.}\xspace}
\newcommand{\eg}{{\em e.g.}\xspace}
\newcommand{\etal}{{\em et al.}\xspace}

\algnewcommand\algorithmicswitch{\textbf{switch}}
\algnewcommand\algorithmiccase{\textbf{case}}
\algnewcommand\algorithmicassert{\texttt{assert}}
\algnewcommand\Assert[1]{\State \algorithmicassert(#1)}%
\algdef{SE}[SWITCH]{Switch}{EndSwitch}[1]{\algorithmicswitch\ #1\ \algorithmicdo}{\algorithmicend\ \algorithmicswitch}%
\algdef{SE}[CASE]{Case}{EndCase}[1]{\algorithmiccase\ #1}{\algorithmicend\ \algorithmiccase}%
\algtext*{EndSwitch}%
\algtext*{EndCase}%

\newcommand{\PRAGMATIC}{PRAgMaTIc\xspace}
\newcommand{\INTEL}{Intel\textsuperscript\textregistered}
\newcommand{\INTELXEON}{Intel\textsuperscript\textregistered 
Xeon\textsuperscript\textregistered\xspace}
\newcommand{\INTELXEONPHI}{Intel\textsuperscript\textregistered Xeon 
Phi\textsuperscript\texttrademark\xspace}
\newcommand{\RHEL}{Red Hat\textsuperscript\textregistered Enterprise 
Linux\textsuperscript\textregistered\xspace}

\newcommand{\StatexIndent}[1][3]{%
  \setlength\@tempdima{\algorithmicindent}%
  \Statex\hskip\dimexpr#1\@tempdima\relax}
\algdef{S}[IF]{IfNoThen}[1]{\algorithmicif\ #1}%

\definecolor{listinggray}{gray}{0.9}
\definecolor{lbcolor}{rgb}{0.9,0.9,0.9}
\floatstyle{plain}
\newfloat{code}{h}{}
\floatname{code}{Code Snippet}

\title{A Fast and Scalable Graph Coloring Algorithm for Multi-core and 
Many-core Architectures}

\author{Georgios Rokos\inst{1} \and Gerard Gorman\inst{2} \and Paul H J 
Kelly\inst{1}}

\institute{Software Peroformance Optimisation Group,\\ Department of 
Computing \and
Applied Modelling and Computation Group,\\
Department of Earth Science and Engineering \\
Imperial College London,\\ South Kensington Campus, London SW7 2AZ, United 
Kingdom,\\
\email{\{georgios.rokos09,g.gorman,p.kelly\}@imperial.ac.uk}}

\begin{document}

\maketitle

\begin{abstract}
Irregular computations on unstructured data are an important class of problems 
for parallel programming. Graph coloring is often an important preprocessing 
step, e.g. as a way to perform dependency analysis for safe parallel execution. 
The total run time of a coloring algorithm adds to the overall parallel 
overhead of the application whereas the number of colors used determines the 
amount of exposed parallelism. A fast and scalable coloring algorithm using as 
few colors as possible is vital for the overall parallel performance and 
scalability of many irregular applications that depend upon runtime dependency 
analysis.

\c{C}ataly{\"u}rek et al. have proposed a graph coloring algorithm which relies 
on speculative, local assignment of colors. In this paper we present an 
improved version which runs even more optimistically with less thread 
synchronization and reduced number of conflicts compared to \c{C}ataly{\"u}rek 
et al.'s algorithm. We show that the new technique scales better on multi-core 
and many-core systems and performs up to 1.5x faster than its predecessor on 
graphs with high-degree vertices, while keeping the number of colors at the 
same near-optimal levels.


\end{abstract}

\section{Introduction}
Many modern applications are built around algorithms which operate on irregular 
data structures, usually in form of graphs. Graph coloring is an important 
preprocessing step, mainly as a means of guaranteeing safe parallel execution 
in a shared-memory environment but also in order to enforce neighborhood 
heuristics, \eg avoid having adjacent graph edges collapse in sequence in graph 
coarsening \cite{de1999parallel}. Examples of such applications include 
iterative methods for sparse linear systems 
\cite{Jones:1994:SIS:180106.180110}, sparse tiling 
\cite{StroutLCPC2002,Strout14IPDPS}, eigenvalue computation 
\cite{Manne:1998:PAC:645781.666669}, preconditioners
\cite{doi:10.1137/0917054,Hysom:2000:SPA:587165.587404} and mesh adaptivity 
\cite{Freitag98thescalability,GormanRSK14}.

Taking advantage of modern multi-core and many-core hardware requires not only 
algorithmic modifications to deal with data races but also consideration of 
scalability issues. The exposed parallelism of an irregular algorithm is 
directly dependent on the number of colors used. The lower this number, the 
more work-items are available for concurrent processing per color/independent 
set. Additionally, there is usually some thread synchronization or reduction 
before proceeding to the next independent set. A poor-quality coloring will 
only exaggerate the effects of thread synchronization on the parallel 
scalability of an application. Following this observation, it is obvious that a 
good coloring algorithm should be fast and scalable itself, so as to minimize 
its own contribution to the total execution time of the application, and use as 
few colors as possible.

The simplest graph coloring algorithm is the greedy one, commonly known as {\em 
First-Fit} (\S \ref{subsect:first_fit}). There exist parallel versions for 
distributed-memory environments, but in this paper we focus on the intra-node, 
shared-memory case. Probably, the best known parallel algorithm is the one by 
Jones and Plassmann \cite{Jones92aparallel}, which in turn is an improved 
version of the original {\em Maximal Independent Set} algorithm by Luby 
\cite{Luby:1985:SPA:22145.22146}. There also exists a modified version of 
Jones-Plassmann which uses multiple hashes to minimize thread synchronization 
\cite{cohen_castonguay}. A parallel greedy coloring algorithm based on 
speculative execution was introduced by Gebremedhin and Manne 
\cite{CPE:CPE528}. \c{C}ataly\"{u}rek \etal presented an improved version 
of the original speculative algorithm in \cite{ccatalyurek2012graph} 
(\S\ref{subsect:catalyurek}). We took the latter one step further, devising a 
method which runs under an even more speculative scheme with less thread 
synchronization (\S\ref{sect:implementation}), without compromising coloring 
quality.

It must be pointed out that First-Fit variants which use ordering heuristics 
were not considered here. Despite recent innovations by Hasenplaugh \etal 
\cite{Hasenplaugh:2014:OHP:2612669.2612697}, those variants take considerably 
longer to run than the plain greedy algorithm and in many cases do not achieve 
a sufficiently large improvement in the number of colors to justify their cost. 
Runtime of coloring for the purpose of dynamic dependency analysis becomes a 
serious consideration in problems like morph algorithms \cite{nasre13-ppopp}, 
which mutate graph topology in non-trivial ways and constantly invalidate 
existing colorings. In those cases, the graph has to be recolored in every 
iteration of the morph kernel, so coloring becomes a recurring cost rather than 
a one-off preprocessing step. As shown in 
\cite{Hasenplaugh:2014:OHP:2612669.2612697}, heuristic-based algorithms, 
although achieving some reduction in the number of colors, take 4x-11x longer 
to run and this would dominate the kernel's runtime. A notable example is the 
edge-swap kernel from our mesh adaptivity framework \PRAGMATIC 
\footnote{https://github.com/meshadaptation/pragmatic} \cite{GormanRSK14}, in 
which coloring (using our fast method) already takes up 10\% of the total 
execution time.

The rest of this paper is organized as follows: In Section 
\ref{sect:background} we present the serial greedy coloring algorithm and its 
parellel implementation by \c{C}ataly\"{u}rek \etal. We explain how the latter 
can be improved further, leading to our implementation which is described in 
Section \ref{sect:implementation} and evaluated against its predecessor in 
Section \ref{sect:coloring_results}. Finally, we briefly explain why the class 
of optimistic coloring algorithms is unsuitable for SIMT-style parallel 
processing systems in Section \ref{sect:simt_coloring} and conclude the paper 
in Section \ref{sect:conclusions}.

\section{Background}
\label{sect:background}
In this section we describe the greedy coloring algorithm and its parallel 
version proposed by \c{C}ataly\"{u}rek \etal.

\subsection{First-Fit Coloring}
\label{subsect:first_fit}
Coloring a graph with the minimal number of colors has been shown to be an 
NP-hard problem \cite{GJ79}. However, there exist heuristic algorithms which 
color a graph in polynomial time using relatively few colors, albeit not 
guaranteeing an optimal coloring. One of the most common polynomial coloring 
algorithms is {\em First-Fit}, also known as {\em greedy coloring}. In its 
sequential form, First-Fit visits every vertex and assigns the smallest color 
available, \ie not already assigned to one of the vertex's neighbors. The 
procedure is summarized in Algorithm \ref{alg:greedy}.

\begin{algorithm}
  \caption{Sequential greedy coloring algorithm.}
  \label{alg:greedy}
  \begin{algorithmic}
	\State Input: $\mathcal{G}(V,E)$
	\ForAll{vertices $V_i \in \mathcal{G}$}
	  \State $\mathcal{C} \gets $ \{colors of all colored vertices $V_j \in 
	  adj(V_i)$\}
	  \State $c(V_i) \gets $ \{smallest color $\not \in \mathcal{C}$\}
	\EndFor
  \end{algorithmic}
\end{algorithm}

It is easy to give an upper bound on the number of colors used by the greedy 
algorithm. Let us assume that the highest-degree vertex $V_h$ in a graph has 
degree $d$, \ie this vertex has $d$ neighbors. In the worst case, each 
neighbor has been assigned a unique color; then one of the colors 
\{$1,2,\dots,d+1$\} will be available to $V_h$ (\ie not already assigned to a 
neighbor). Therefore, the greedy algorithm can color a graph with at most 
$d+1$ colors. In fact, experiments have shown that First-Fit can produce 
near-optimal colorings for many classes of graphs \cite{doi:10.1137/0720013}.

\subsection{Optimistic Coloring}
\label{subsect:catalyurek}
Gebremedhin and Manne introduced an optimistic approach to parallelizing the 
greedy graph coloring algorithm \cite{CPE:CPE528}. They described a fast and 
scalable version for shared-memory systems based on the principles of 
speculative (or optimistic) execution. The idea is that we can color all 
vertices in parallel using First-Fit without caring about race conditions at 
first (stage 1); this can lead to defective coloring, \ie two adjacent vertices 
might get the same color. Defects can then be spotted in parallel (stage 2) and 
fixed by a single thread (stage 3).

Picking up where Gebremedhin and Manne left off, \c{C}ataly\"{u}rek \etal 
improved the original algorithm by removing the sequential conflict-resolution 
stage and applying the first two parallel stages iteratively. This work was 
presented in \cite{ccatalyurek2012graph}. Each of the two phases, called {\em 
tentative coloring} phase and {\em conflict detection} phase respectively, is 
executed in parallel over a relevant set of vertices. Like the original 
algorithm by Gebremedhin and Manne, the tentative coloring phase produces a 
pseudo-coloring of the graph, whereas in the conflict detection phase threads 
identify defectively colored vertices and append them into a list 
$\mathcal{L}$. Instead of resolving conflicts in $\mathcal{L}$ serially, 
$\mathcal{L}$ now forms the new set of vertices over which the next execution 
of the tentative coloring phase will iterate. This process is repeated until 
no conflicts are encountered.

\begin{algorithm}
  \caption{The parallel graph coloring algorithm by \c{C}ataly\"{u}rek \etal.}
  \label{alg:catalyurek}
  \begin{algorithmic}
    \State Input: $\mathcal{G}(V,E)$
    \State $\mathcal{U} \gets V$
    \While{$\mathcal{U} \not = \emptyset$}
      \State \textbf{\#pragma omp parallel for} \Comment Phase 1 - Tentative 
      coloring (in parallel)
	  \ForAll{vertices $V_i \in \mathcal{U}$} \Comment execute First-Fit
	    \State $\mathcal{C} \gets $ \{colors of all colored vertices $V_j \in 
	    adj(V_i)$\}
	    \State $c(V_i) \gets $ \{smallest color $\not \in \mathcal{C}$\}
	  \EndFor   
	  \State \textbf{\#pragma omp barrier}
	  \State $\mathcal{L} \gets \emptyset$ \Comment global list of defectively 
	  colored vertices
	  \State \textbf{\#pragma omp parallel for} \Comment Phase 2 - Conflict 
	  detection (in parallel)
	  \ForAll{vertices $V_i \in \mathcal{U}$}
	    \If{$\exists V_j \in adj(V_i), V_j > V_i: c(V_j)==c(V_i)$}
	      \State $\mathcal{L} \gets \mathcal{L} \cup V_i$ \Comment mark $V_i$ 
	      as defectively colored
	    \EndIf
	  \EndFor
	  \State \textbf{\#pragma omp barrier}
	  \State $\mathcal{U} \gets \mathcal{L}$ \Comment Vertices to be re-colored 
	  in the next round
    \EndWhile	
  \end{algorithmic}
\end{algorithm}

Algorithm \ref{alg:catalyurek} summarizes this coloring method. As can be 
seen, there is no sequential part in the whole process. Additionally, speed 
does not come at the expense of coloring quality. The authors have demonstrated 
that this algorithm produces colorings using about the same number of colors as 
the serial greedy algorithm. However, there is still a source of sequentiality, 
namely the two thread synchronization points in every iteration of the 
while-loop. Synchronization can easily become a scalability barrier for high 
numbers of threads and should be minimized or eliminated if possible.

\section{Implementation}
\label{sect:implementation}
Moving toward the direction of removing as much thread synchronization as 
possible, we improved the algorithm by \c{C}ataly\"{u}rek \etal by eliminating 
one of the two barriers inside the while-loop. This was achieved by merging the 
two parallel for-loops into a single parallel for-loop. We observed that when a 
vertex is found to be defective it can be re-colored immediately instead of 
deferring its re-coloring for the next round. Therefore, the tentative-coloring 
and conflict-detection phases can be combined into a single {\em 
detect-and-recolor} phase in which we inspect all vertices which were 
re-colored in the previous iteration of the while-loop. Doing so leaves only 
one thread synchronization point per round, as can be seen in Algorithm 
\ref{alg:rsoc}. This barrier guarantees that any changes committed by a thread 
are made visible system-wide before proceeding to the next round.

\begin{algorithm}
  \caption{The improved parallel graph coloring technique.}
  \label{alg:rsoc}
  \begin{algorithmic}
    \State Input: $\mathcal{G}(V,E)$
    \State \textbf{\#pragma omp parallel for} \Comment perform tentative 
    coloring on $\mathcal{G}$; round 0
    \ForAll{vertices $V_i \in \mathcal{G}$}
      \State $\mathcal{C} \gets $ \{colors of all colored vertices $V_j \in 
      adj(V_i)$\}
      \State $c(V_i) \gets $ \{smallest color $\not \in \mathcal{C}$\}
    \EndFor
    \State \textbf{\#pragma omp barrier}
    \State $\mathcal{U}^0 \gets V$ \Comment mark all vertices for inspection
    \State $i \gets 1$ \Comment round counter
    \While{$\mathcal{U}^{i-1} \not = \emptyset$} \Comment $\exists$ vertices 
    (re-)colored in the last round
      \State $\mathcal{L} \gets \emptyset$ \Comment global list of defectively 
      colored vertices
      \State \textbf{\#pragma omp parallel for}
	  \ForAll{vertices $V_i \in \mathcal{U}^{i-1}$}
	    \If{$\exists V_j \in adj(V_i), V_j > V_i: c(V_j)==c(V_i)$} \Comment if 
	    they are (still) defective
	      \State $\mathcal{C} \gets $ \{colors of all colored $V_j \in 
	      adj(V_i)$\} \Comment re-color them
	      \State $c(V_i) \gets $ \{smallest color $\not \in \mathcal{C}$\}
	      \State $\mathcal{L} \gets \mathcal{L} \cup V_i$ \Comment $V_i$ was 
	      re-colored in this round
	  	\EndIf
	  \EndFor
	  \State \textbf{\#pragma omp barrier}
	  \State $\mathcal{U}_{i} \gets \mathcal{L}$ \Comment Vertices to be 
	  inspected in the next round
	  \State $i \gets i+1$ \Comment proceed to the next round
    \EndWhile	
  \end{algorithmic}
\end{algorithm}

\section{Experimental Results}
\label{sect:coloring_results}
In order to evaluate our improved coloring method, henceforth referred to 
as {\em Reduced Synchronization Optimistic Coloring} (RSOC), and compare it to 
the previous state-of-the-art technique by \c{C}ataly\"{u}rek \etal, we ran a 
series of benchmarks using 2D and 3D meshes of triangular and tetrahedral 
elements respectively (commonly used in finite element and finite volume 
methods), alongside randomly generated graphs using the {\em R-MAT} graph 
generation algorithm \cite{Chakrabarti06graphmining}. Simplicial 2D/3D meshes 
are used in order to measure performance and scalability for our target 
application area (\cite{GormanRSK14}), whereas RMAT graphs were used for 
consistency with the experimental methodology used in \c{C}ataly\"{u}rek 
\etal's publication; the authors state that those RMAT graphs ``are designed to 
represent instances posing varying levels of difficulty for the performance of 
multithreaded coloring algorithms'' \cite{ccatalyurek2012graph}.

For the 2D case we have used a 2D anisotropic mesh (adapted to the requirements 
of some CFD problem) named \verb=mesh2d=, which consists of $\approx 250k$ 
vertices. We also evaluate performance using two 3D meshes, taken from the 
University of Florida Sparse Matrix Collection 
\cite{Davis:2011:UFS:2049662.2049663}. \verb=bmw3_2= is a mesh modelling a BMW 
Series 3 car consisting of $\approx 227k$ vertices, whereas \verb=pwtk= 
represents a pressurized wind tunnel and consists of $\approx 218k$ vertices. 
Finally, we generated three $16M$-vertex, $128M$-edge RMAT graphs, namely 
\verb=RMAT-ER= (Erd\H{o}s-R\'{e}nyi), \verb=RMAT-G= (Good) and \verb=RMAT-B= 
(Bad), randomly shuffling vertex indices so as to reduce the benefits of data 
locality and large caches. For more information on those graphs the reader is 
referred to the original publication by \c{C}ataly\"{u}rek \etal 
\cite{ccatalyurek2012graph}.

The experiments were run on two systems: a dual-socket \INTELXEON E5-2650 
system (Sandy Bridge, 2.00GHz, 8 physical cores per socket, 2-way 
hyper-threading) running \RHEL Server release 6.4 (Santiago) and an 
\INTELXEONPHI 5110P board (1.053GHz, 60 physical cores, 4-way hyper-threading). 
Both versions of the code (intel64 and mic) were compiled with \INTEL Composer 
XE 2013 SP1 and with the compiler flags {\tt -O3 -xAVX}. The benchmarks were 
run using \INTEL's thread-core affinity support.

Table \ref{tab:times} shows the average execution time over 10 runs of both 
algorithms on the 2 systems, \INTELXEON and \INTELXEONPHI, using the 3 finite 
element/volume meshes and the 3 RMAT graphs. Rows preceded by ``C'' correspond 
to the algorithm by \c{C}ataly\"{u}rek \etal, rows preceded by ``R'' pertain to 
the improved version. Timings for the meshes are given in milliseconds whereas 
for the RMAT graphs they are in seconds. As can be seen, RSOC performs faster 
than \c{C}ataly\"{u}rek \etal for every test graph on both platforms, while 
scaling better as the number of threads increases, especially on \INTELXEONPHI.

\begin{table}[!h]
\begin{center}
\begin{scriptsize}
\begin{tabular}[c]{|lc|c|c|c|c|c|c||c|c|c|c|c|c|c|c|c|}
\cline{3-17}
\multicolumn{2}{c|}{}				& \multicolumn{6}{|c||}{\multirow{2}{*}{\INTELXEON}}	& \multicolumn{9}{c|}{\multirow{2}{*}{\INTELXEONPHI}}					\\
\multicolumn{2}{c|}{} 				& \multicolumn{6}{|c||}{}								& \multicolumn{9}{c|}{}													\\ \cline{3-17}
\multicolumn{2}{c|}{}				& \multicolumn{6}{|c||}{Number of OpenMP threads}		& \multicolumn{9}{c|}{Number of OpenMP threads}							\\
\multicolumn{2}{c|}{}				& 1		& 2		& 4		& 8		& 16	& 32 			& 1		& 2		& 4		& 8		& 15	& 30	& 60	& 120	& 240	\\ \hline
\multirow{2}{*}{mesh2d}		& C:	& 62.7	& 34.0	& 19.2	& 10.2	& 5.92	& 4.28			& 496	& 252	& 127	& 64.9	& 35.5	& 19.0	& 11.7	& 12.7	& 73.6	\\
							& R:	& 62.2	& 31.3	& 17.7	& 9.42	& 5.50	& 4.05			& 495	& 249	& 125	& 63.3	& 34.5	& 17.9	& 10.7	& 10.5	& 69.4	\\ \hline
\multirow{2}{*}{bmw3\_2}	& C:	& 58.1	& 33.5	& 14.4	& 7.84	& 4.73	& 3.61			& 468	& 235	& 118	& 60.0	& 33.1	& 18.0	& 11.5	& 12.7	& 74.2	\\
							& R:	& 57.8	& 29.4	& 12.1	& 6.48	& 3.91	& 3.30			& 466	& 234	& 117	& 59.2	& 32.4	& 17.1	& 9.88	& 11.0	& 54.9	\\ \hline
\multirow{2}{*}{pwtk}		& C:	& 40.1	& 24.0	& 14.5	& 8.07	& 4.96	& 3.65			& 465	& 233	& 117	& 59.6	& 33.2	& 18.2	& 11.1	& 12.9	& 74.4	\\
							& R:	& 39.8	& 20.0	& 11.3	& 6.08	& 3.81	& 3.30			& 464	& 232	& 117	& 58.9	& 32.4	& 17.2	& 10.6	& 11.0	& 59.9	\\ \hline \hline
\multirow{2}{*}{RMAT-ER}	& C:	& 6.11	& 3.21	& 1.82	& 1.09	& 0.79	& 0.85			& 196	& 97.8	& 48.9	& 24.6	& 13.0	& 6.41	& 3.16	& 1.64	& 0.94	\\
							& R:	& 6.09	& 3.20	& 1.81	& 1.08	& 0.78	& 0.85			& 196	& 98.0	& 49.0	& 24.7	& 13.1	& 6.43	& 3.16	& 1.64	& 0.95	\\ \hline
\multirow{2}{*}{RMAT-G}		& C:	& 6.10	& 3.18	& 1.82	& 1.08	& 0.77	& 0.81			& 195	& 97.1	& 48.6	& 24.3	& 12.9	& 6.34	& 3.12	& 1.62	& 0.93	\\
							& R:	& 6.07	& 3.17	& 1.81	& 1.07	& 0.77	& 0.81			& 195	& 97.3	& 48.7	& 24.4	& 13.0	& 6.38	& 3.13	& 1.63	& 0.93	\\ \hline
\multirow{2}{*}{RMAT-B}		& C:	& 5.47	& 2.86	& 1.62	& 0.93	& 0.65	& 0.64			& 189	& 94.1	& 46.7	& 23.5	& 12.3	& 6.08	& 3.12	& 1.90	& 1.49	\\
							& R:	& 5.46	& 2.83	& 1.60	& 0.92	& 0.64	& 0.63			& 189	& 94.0	& 46.9	& 23.5	& 12.4	& 6.02	& 2.95	& 1.60	& 1.00	\\ \hline
\end{tabular}
\end{scriptsize}

\caption{Execution time of both algorithms on 2 different platforms, \INTELXEON 
and \INTELXEONPHI, with varying number of OpenMP threads and using the 3 finite 
element/volume meshes and the 3 RMAT graphs. Rows preceded by ``C'' correspond 
to the algorithm by \c{C}ataly\"{u}rek \etal, rows preceded by ``R'' pertain to 
the improved version. Timings for the meshes are given in milliseconds whereas 
for the graphs they are in seconds.}
\label{tab:times}
\end{center}
\end{table}

\begin{figure}
\centering
\includegraphics[scale=0.85]{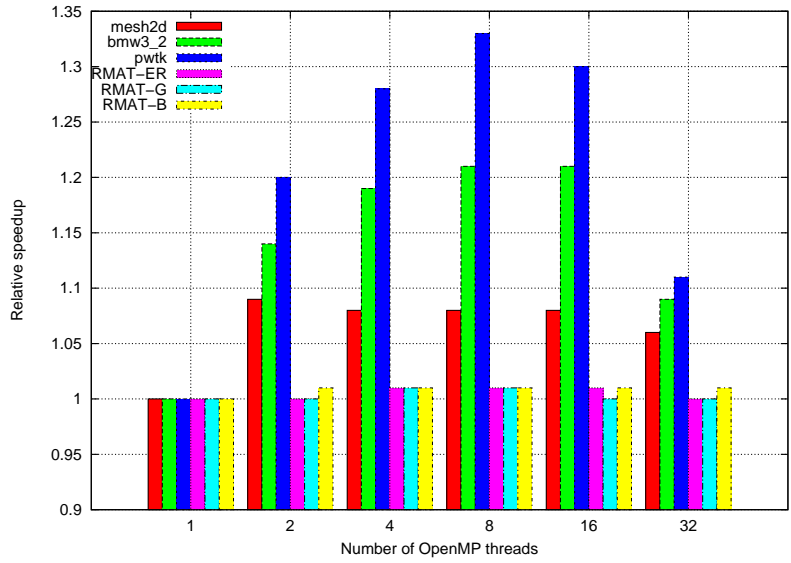}
\caption{Speedup of RSOC relative to \c{C}ataly\"{u}rek \etal as the number of 
threads increases on \INTELXEON E5-2650.}
\label{fig:rel_speedup_host}
\end{figure}

\begin{figure}
\centering
\includegraphics[scale=0.85]{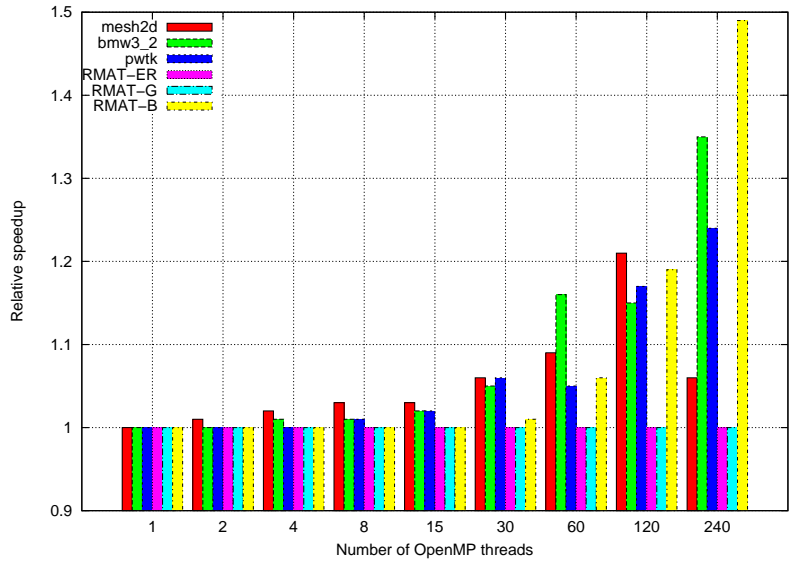}
\caption{Speedup of RSOC relative to \c{C}ataly\"{u}rek \etal as the number of 
threads increases on \INTELXEONPHI 5110P.}
\label{fig:rel_speedup_phi}
\end{figure}

Figures \ref{fig:rel_speedup_host} and \ref{fig:rel_speedup_phi} show the 
relative speedup of RSOC over \c{C}ataly\"{u}rek \etal for all test graphs on 
\INTELXEON and \INTELXEONPHI, respectively, \ie how much faster our 
implementation is than its predecessor for a given number of threads. With the 
exception of \verb=RMAT-ER= and \verb=RMAT-G= on which there is no difference 
in performance, the gap between the two algorithms widens as the number of 
threads increases, reaching a maximum value of 50\% on \INTELXEONPHI for 
\verb=RMAT-B=.

\begin{figure}
\centering
\includegraphics[scale=0.85]{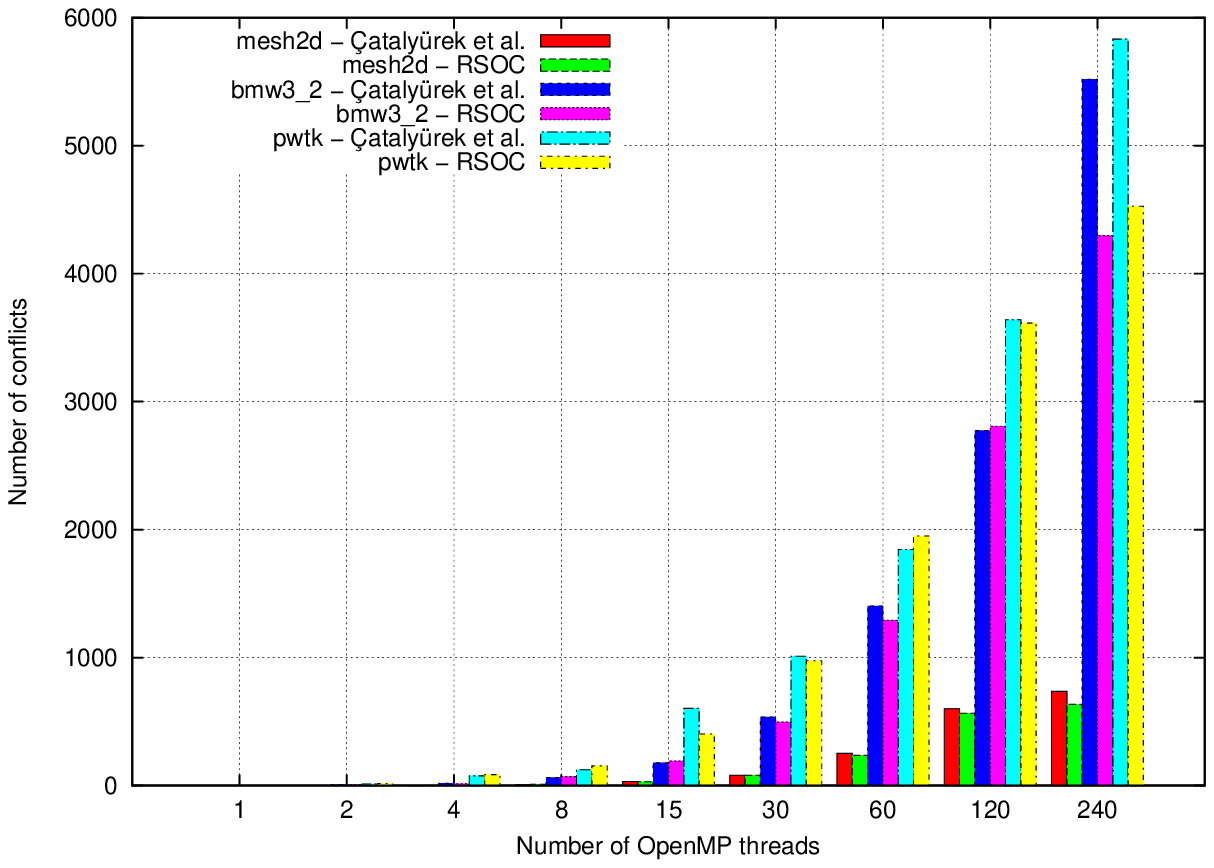}
\cprotect\caption{Number of conflicts on \INTELXEONPHI 5110P using 
\verb=mesh2d=, \verb=bmw3_2= and \verb=pwtk=.}
\label{fig:conflicts_mesh_phi}
\end{figure}

\begin{figure}
\centering
\includegraphics[scale=0.85]{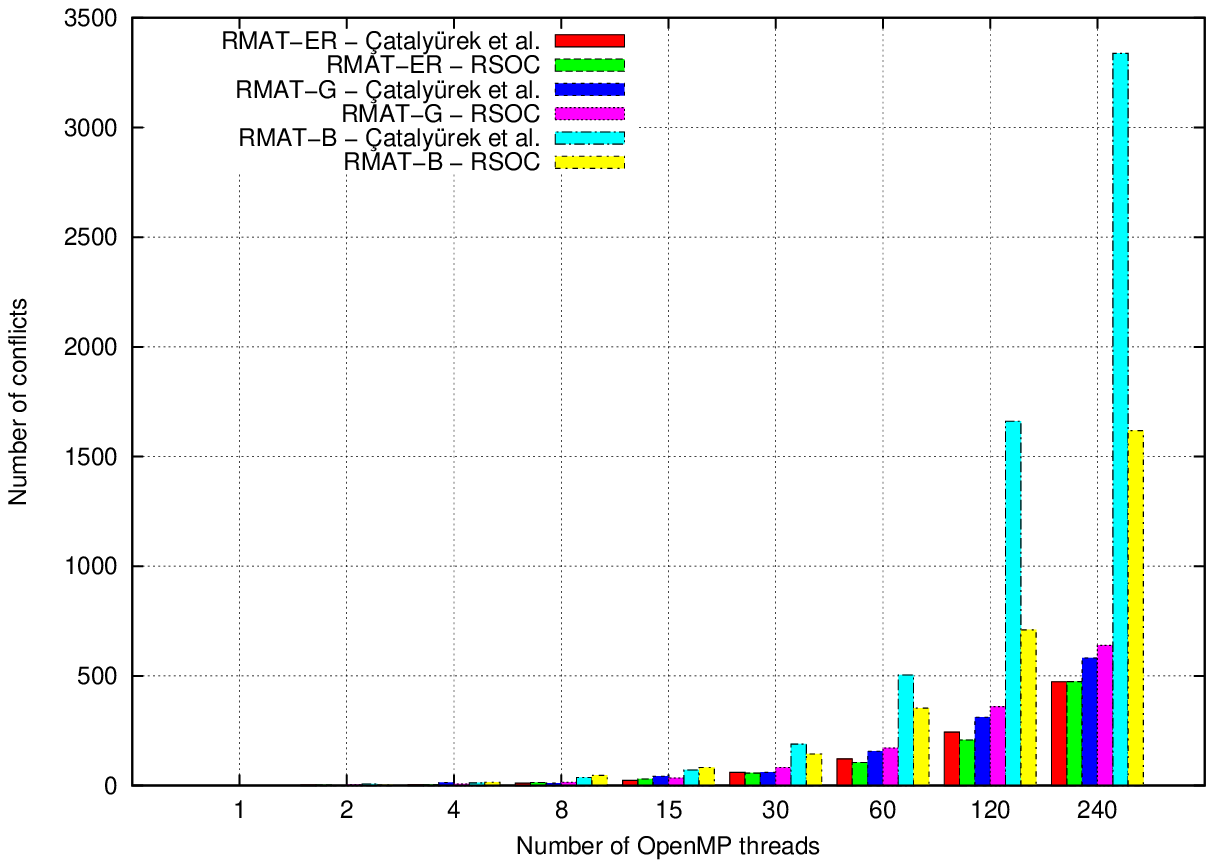}
\cprotect\caption{Number of conflicts on \INTELXEONPHI 5110P using 
\verb=RMAT-ER=, \verb=RMAT-G= and \verb=RMAT-B=.}
\label{fig:conflicts_rmat_phi}
\end{figure}

Looking at the total number of coloring conflicts encountered throughout the 
execution of both algorithms as well as the number of iterations each algorithm 
needs in order to resolve them, we can identify an additional source of speedup 
for our algorithm (apart from the absence of one barrier). We will use the 
\INTELXEONPHI system for this study, as it is the platform on which the most 
interesting results have been observed. Figures \ref{fig:conflicts_mesh_phi} 
and \ref{fig:conflicts_rmat_phi} depict the total number of conflicts for the 
three meshes and the RMAT graphs, respectively. When using few threads both 
algorithms produce about the same number of conflicts. However, moving to 
higher levels of parallelism reveals that RSOC results in much fewer defects in 
coloring for certain classes of graphs.

This observation can be explained as follows: In \c{C}ataly\"{u}rek \etal all 
threads synchronize before entering the conflict-resolution phase, which means 
that they enter that phase and start resolving conflicts at the very same time. 
Therefore, it is highly possible that two adjacent vertices with conflicting 
colors will be processed by two threads simultaneously, which leads once again 
to new defects. In our improved algorithm, on the other hand, a conflict is 
resolved as soon as it is discovered by a thread. The likelihood that another 
thread is recoloring a neighboring vertex at the same time is certainly lower 
than in \c{C}ataly\"{u}rek \etal.

%

\begin{figure}
\centering
\includegraphics[scale=0.85]{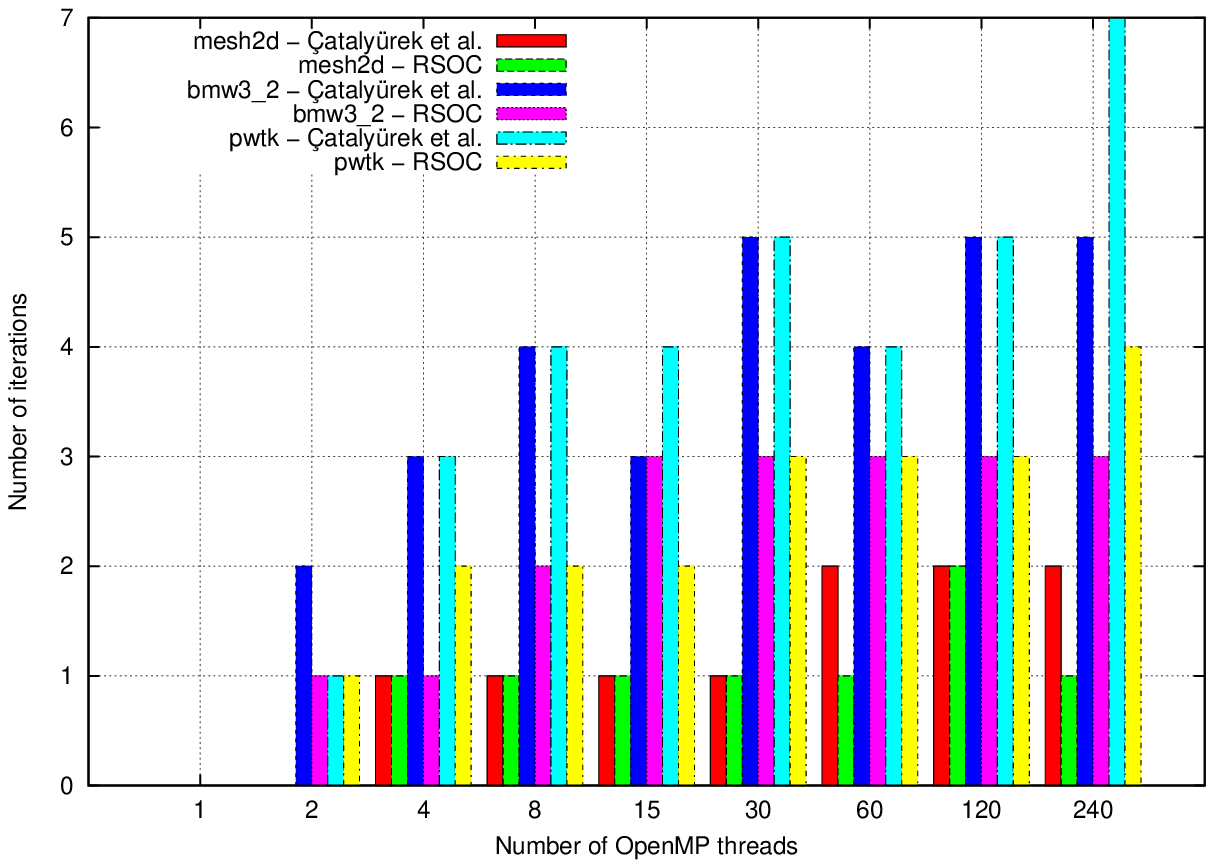}
\cprotect\caption{Number of iterations on \INTELXEONPHI 5110P using 
\verb=mesh2d=, \verb=bmw3_2= and \verb=pwtk=.}
\label{fig:iterations_mesh_phi}
\end{figure}

\begin{figure}
\centering
\includegraphics[scale=0.85]{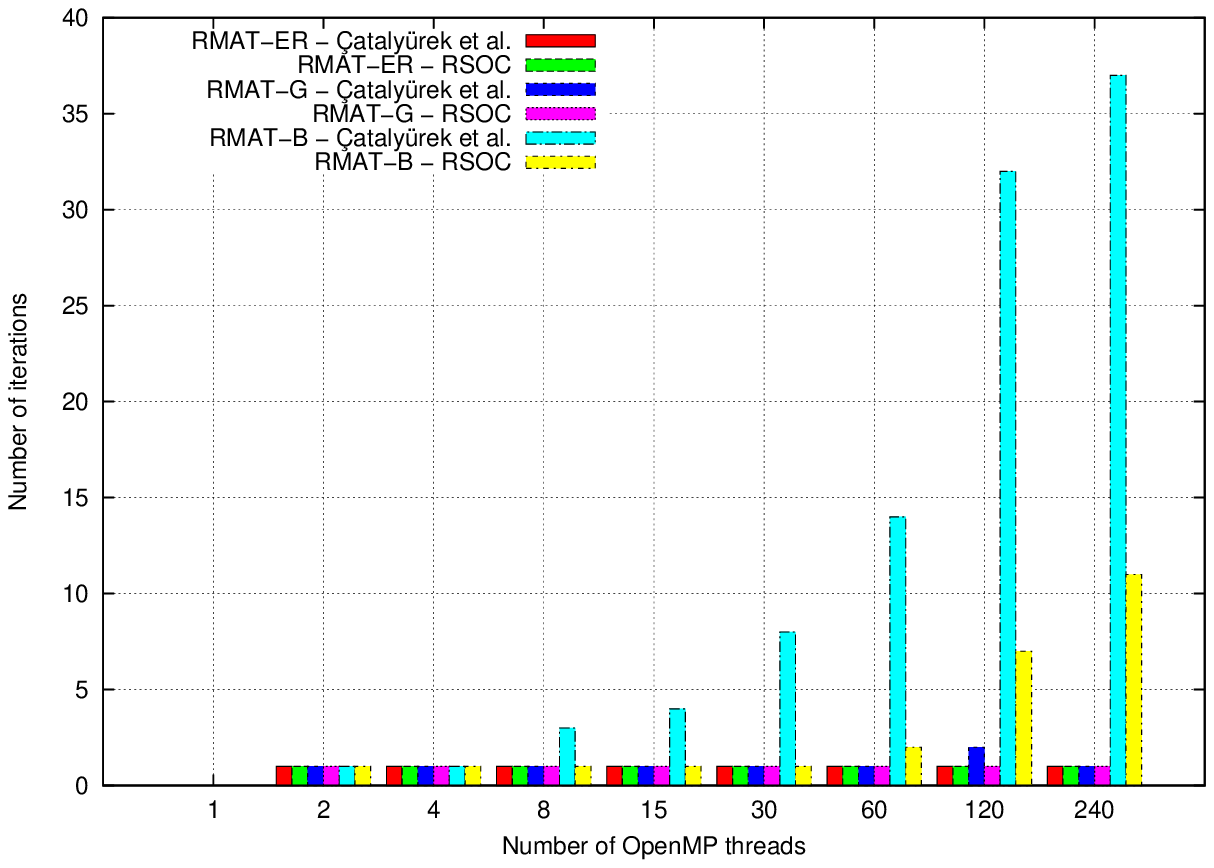}
\cprotect\caption{Number of iterations on \INTELXEONPHI 5110P using 
\verb=RMAT-ER=, \verb=RMAT-G= and \verb=RMAT-B=.}
\label{fig:iterations_rmat_phi}
\end{figure}

The reduced number of conflicts also results in fewer iterations of the 
algorithm, as can be seen in Figures \ref{fig:iterations_mesh_phi} and 
\ref{fig:iterations_rmat_phi}. Combined with the absence of one barrier from 
the while-loop, it is only expected that our new algorithm ultimately 
outperforms its predecessor. A nice property is that both algorithms produce 
colorings using the same number of colors, \ie quality of coloring is not 
compromised by the higher execution speed.

%

\section{SIMT restrictions}
\label{sect:simt_coloring}
\begin{figure}
\begin{minipage}{0.24\linewidth}
\begin{center}
\begin{tikzpicture}[scale=1]
\node[draw,shape=circle] (a) at ( 0, 0) {0};
\node[draw,shape=circle] (b) at ( 1, 0) {1};
\draw [-] (a) -- (b);
\end{tikzpicture}
\end{center}
\begin{center}
(a) Graph
\end{center}
\end{minipage}
\begin{minipage}{0.24\linewidth}
\begin{center}
\begin{tikzpicture}[scale=1]
\node[draw,shape=circle,fill=red] (a) at ( 0, 0) {0};
\node[draw,shape=circle,fill=red] (b) at ( 1, 0) {1};
\draw [-] (a) -- (b);
\end{tikzpicture}
\end{center}
\begin{center}
(b) Round 1
\end{center}
\end{minipage}
\begin{minipage}{0.24\linewidth}
\begin{center}
\begin{tikzpicture}[scale=1]
\node[draw,shape=circle,fill=green] (a) at ( 0, 0) {0};
\node[draw,shape=circle,fill=green] (b) at ( 1, 0) {1};
\draw [-] (a) -- (b);
\end{tikzpicture}
\end{center}
\begin{center}
(c) Round 2
\end{center}
\end{minipage}
\begin{minipage}{0.24\linewidth}
\begin{center}
\begin{tikzpicture}[scale=1]
\node[draw,shape=circle,fill=red] (a) at ( 0, 0) {0};
\node[draw,shape=circle,fill=red] (b) at ( 1, 0) {1};
\draw [-] (a) -- (b);
\end{tikzpicture}
\end{center}
\begin{center}
(d) Round 3
\end{center}
\end{minipage}
\caption{Example of an infinite loop in SIMT-style parallelism when using one 
of the optimistic coloring algorithms.}
\label{fig:simt_coloring}
\end{figure}
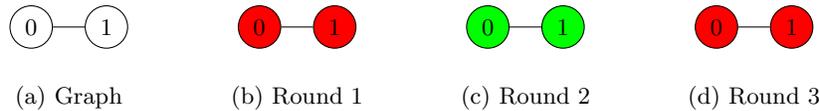

Trying to run the optimistic coloring algorithms using CUDA on an Nvidia GPU 
revealed a potential weakness. Neither algorithm terminated; instead, threads 
spun forever in an infinite loop. This is due to the nature of SIMT-style 
multi-threading, in which the lockstep warp execution results in ties never 
being broken. An example of why these algorithms result in infinite loops in 
SIMT-style parallelism can be seen in Figure \ref{fig:simt_coloring}, where we 
have a simple two-vertex graph and two threads, each processing one vertex 
(this scenario is likely to actually occur at a later iteration of the 
while-loop, where the global list of defects $\mathcal{L}$ is left with a few 
pairs of adjacent vertices). At the beginning (a), both vertices are 
uncolored. Each thread decides that the smallest color available for its own 
vertex is red. Both threads commit their decision at the same clock cycle, 
which results in the defective coloring shown in (b). In the next round the 
threads try to resolve the conflict and decide that the new smallest color 
available is green. The decision is committed at the same clock cycle, 
resulting once again in defects (c) and the process goes on forever.

Theoretically, this scenario is possible for CPUs as well, although the 
probability is extremely low. We believe that there will always be some 
randomness (\ie lack of thread coordination) on CPUs which guarantees 
convergence of the optimistic algorithms. This randomness can also be 
``emulated'' on GPUs by having a dynamic assignment of vertices to threads and 
making sure that two adjacent vertices are always processed by threads of 
different warps.

\section{Conclusions}
\label{sect:conclusions}
In this article we presented an older parallel graph coloring algorithm and 
showed how we devised an improved version which outperforms its predecessor, 
being up to 50\% faster for certain classes of graphs and scaling better on 
manycore architectures. The difference becomes more pronounced as we move to 
graphs with higher-degree vertices (3D meshes, RMAT-B graph).

This observation also implies that our method (with the appropriate extensions) 
could be a far better option for d-distance colorings of a graph $\mathcal{G}$, 
where $\mathcal{G}^d$ is considerably more densely connected than $\mathcal{G}$ 
(graph $\mathcal{G}^d$, the $d^{th}$ power graph of $\mathcal{G}$, has the same 
vertex set as $\mathcal{G}$ and two vertices in $\mathcal{G}^d$ are connected 
by an edge if and only if the same vertices are within distance $d$ in 
$\mathcal{G}$).

Speed and scalability stem from two sources, (a) reduced number of conflicts 
which also results in fewer iterations and (b) reduced thread synchronization 
per iteration. Coloring quality remains at the same levels as in older 
parallel algorithms, which in turn are very close to the serial greedy 
algorithm, meaning that they produce near-optimal colorings for most classes 
of graphs.

\section*{Acknowledgments}
The authors gratefully acknowledge funding from EPSRC grants EP/I00677X/1 and 
EP/L000407/1 supporting this work.

\bibliographystyle{splncs03}
\bibliography{paper}

\end{document}